%% file: sample.tex
\documentclass{applemlr}

\input{apple_preamble}

\theoremstyle{plain}

\usetikzlibrary{arrows.meta}

\newcommand{\method}{\textsc{CoGR}\xspace}

\definecolor{promptbg}{HTML}{FCF2F2}
\definecolor{promptframe}{HTML}{E3C9C9}
\definecolor{promptred}{HTML}{C0392B}
\newcommand{\ph}[1]{\textcolor{promptred}{\{#1\}}}

\usepackage{fontawesome5}
\newcommand{\frozen}{\textsuperscript{\scriptsize\faSnowflake}}

\usepackage{CJKutf8}
\newcommand{\cjk}[1]{\begin{CJK*}{UTF8}{gbsn}#1\end{CJK*}}

\title{It Takes Two to Match: Co-Evolving Generative Retriever with Reinforcement Learning}

\author[1,2,\ddagger]{Runpeng Dai}
\author[2]{Kaili Huang}
\author[2]{Changsung Kang}
\author[2]{Ciya Liao}
\affiliation{$^{1}$University of North Carolina at Chapel Hill \quad $^{2}$Apple}

\abstract{
Retrieval is the first stage of modern search and advertising systems, selecting a candidate set from a large item universe for downstream ranking and auction. Recent work increasingly leverages LLMs to improve retrieval through query expansion, data synthesis, and retrieval-feedback training. However, the generative component is typically used for query-side augmentation, while final matching is still delegated to a downstream retriever. We introduce \method{}, a retrieval framework that instead trains LLMs to directly construct retrieval representations on both query and item sides. Each generator produces a compact set of keywords, which are matched directly through an inverted index, preserving compatibility with existing keyword-based retrieval infrastructure. \method{} uses a two-stage training pipeline. Supervised fine-tuning first establishes an aligned keyword space, after which co-evolving reinforcement learning alternately optimizes the query- and item-side generators with GRPO against the opposite side's frozen index. Both sides optimize the same query-to-item retrieval $F_1$ objective: the query side receives retrieval $F_1$ directly, while the item side receives a counterfactual marginal reward measuring the change in query-side $F_1$ caused by its generated keywords. Across 10 representative sparse, dense, and generative baselines, \method{} achieves the best performance on both an internal APP Marketplace dataset and the public WANDS benchmark, improving $F_1$ over the strongest baseline by $10.9\%$ and $36.1\%$, respectively. Further analysis shows stable co-evolution and increasingly aligned query--item keyword spaces over training.}
\metadata[Correspondence]{\sffamily
  Runpeng Dai: \url{runpeng@unc.edu};
  Ciya Liao: \url{ciya.liao@apple.com}}

\begin{document}

\maketitle
% ------------------------------------------------------------
% First-page footnotes:
%   - intern affiliation note for Amin
%   - Apple trademark line
% ------------------------------------------------------------
\applefootnote{\textcolor{textgray}{\sffamily%
  $^{\ddagger}$Work done during the internship at Apple.\\
  Apple and the Apple logo are trademarks of Apple Inc., registered in the U.S. and other countries and regions.}}

\section{Introduction}

\begin{figure*}[h!]
  \centering
    \includegraphics[width=\linewidth]{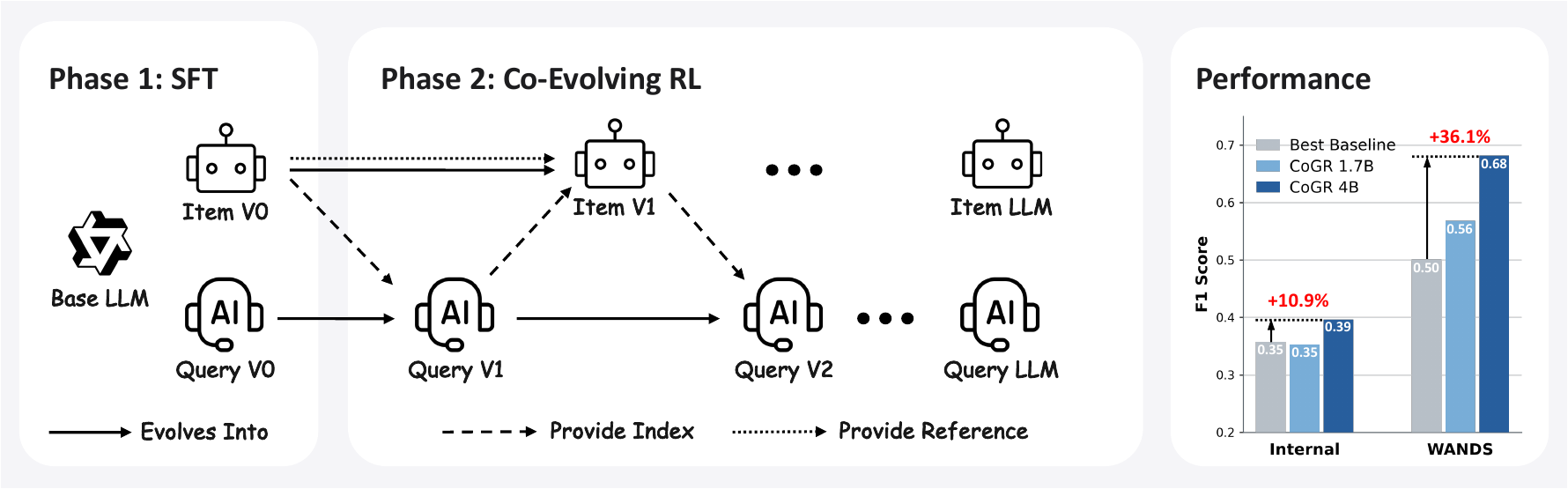}%
  \caption{\textbf{The overall \method{} training pipeline and performance.} \method{} consists a SFT stage and a co-evolving RL stage looping between Item and Query LLMs. \method achieves strong performance gains on both internal APP marketplace and public item search benchmarks.}
  \label{fig:intro}
\end{figure*}

Retrieval is the first stage of modern search and recommendation systems. Given a query, it selects a candidate set from a large item universe for downstream ranking and auction. This stage is consequential because its errors are largely irreversible. An item not retrieved cannot be recovered by later models while irrelevant candidates increase the burden on downstream stages. An effective retriever must therefore maintain broad coverage while keeping the candidate set precise, reflecting the fundamental trade-off between recall and precision \citep{chowdhury2010introduction}.

Classical lexical retrieval methods such as BM25 \citep{robertson2009bm25} match queries and items through explicit terms and inverted indexes. Within this paradigm, keyword-based retrieval is especially prevalent in sponsored search, where advertisers directly bid on keywords. However, lexical representations can be limited in capturing deeper semantic relationships. To move beyond exact lexical overlap, dense retrieval maps queries and items into a shared continuous representation space and matches them through vector similarity \citep{karpukhin2020dpr,xiong2020approximate,ni2022gtr}. More recently, generative retrieval offers another alternative by directly generating semantic identifiers, replacing similarity-based matching with autoregressive prediction \citep{tay2022dsi,wang2022nci,sun2023genret,zeng2024ripor}. However, generative retrieval often depends heavily on identifier design and faces challenges in decoding scalability and generalization \citep{li2025matching}.

Recent work has explored the use of LLMs for retrieval, leveraging their strong capabilities in semantic understanding. Earlier approaches prompt LLMs for query expansion, data synthesis, or keyword generation to improve retrieval \citep{gao2023hyde,wang2023query2doc,ma2023query,liu2025real}. More recent methods further train LLMs using feedback from downstream retrievers \citep{jiang2025deepretrieval,li2025reinforced,yao-etal-2025-expandr}. However, these methods typically train generator of one side, most often the query side and they still rely on a separate downstream retriever for matching. This raises a natural question: can we instead train LLMs to jointly construct retrieval representations for both sides and directly match queries and items in the resulting representation space?

To this end, we propose \method{}, a retrieval framework that trains separate LLMs to generate keywords for queries and items. Given a query or an item, the corresponding generator produces a compact set of keywords, and retrieval is performed directly by matching the generated keyword sets through an inverted index. The generated keywords thus serve directly as retrieval representations. This design also preserves compatibility with existing keyword-based retrieval infrastructure.

The key challenge is to align the two generated keyword spaces so that semantically relevant query--item pairs can be reliably matched. \method{} addresses this with a two-stage training pipeline consisting of supervised fine-tuning (SFT) followed by co-evolving reinforcement learning (RL). The SFT stage establishes an aligned initialization by constructing query-side targets from the keywords of relevant items. Starting from this initialization, we alternately optimize the query- and item-side generators with GRPO \citep{shao2024deepseekmath}. The query-side generator is rewarded directly by the retrieval $F_1$ induced by its generated keywords, while the item-side generator receives a counterfactual marginal reward that measures the change in the same query-side $F_1$ objective caused by replacing its keyword set. During each update, the index produced by the opposite side is kept frozen, allowing each generator to optimize against a fixed retrieval environment while the two keyword spaces progressively co-evolve.

Empirically, we compare \method{} against 10 representative baselines spanning sparse, dense, and generative retrieval. On both the internal APP Marketplace dataset and the public WANDS product-search benchmark \citep{chen2022wands}, \method{} achieves the best overall retrieval performance, improving $F_1$ over the strongest baseline by $10.9\%$ and $36.1\%$, respectively. We further find that alternating query--item optimization is dynamically stable, that co-evolving both sides is important for strong performance, and that the generated keyword spaces become increasingly specific and aligned over training.

%======================================================================
\section{Method}
\label{sec:method}
\subsection{Overview}
\label{sec:overview}
Retrieval is a fundamental task for modern search systems. Given a query $q$, the goal is to retrieve the set of relevant items from a fixed item universe $\mathcal{I}$. Depending on the scenario, the item may be an advertisement, an app, a product, or a document. 

Our method generates keywords for both queries and items and performs keyword-based matching. We train two separate keyword generators: query-side and item-side generators $G^q$, $G^i$. Given a query $q\in \mathcal{Q}$ and an item $i \in \mathcal{I}$, the two generators produce sets of keywords, denoted by $S_q = G^q(q)$ and $S_i = G^i(i)$, respectively. Then we get the retrieved item set $I_{\mathrm{ret}}(q) = \{ i : (S_q \cup \{q\}) \cap (S_i \cup \{i\}) \neq \varnothing \}$, containing items whose keyword sets overlap. To enable retrieval ranking, we treat each side's keyword set as a bag of words and rank the retrieved items by their BM25 score.

\begin{figure*}[t!]
  \centering
  \IfFileExists{figure/main.pdf}{%
    \includegraphics[width=\linewidth]{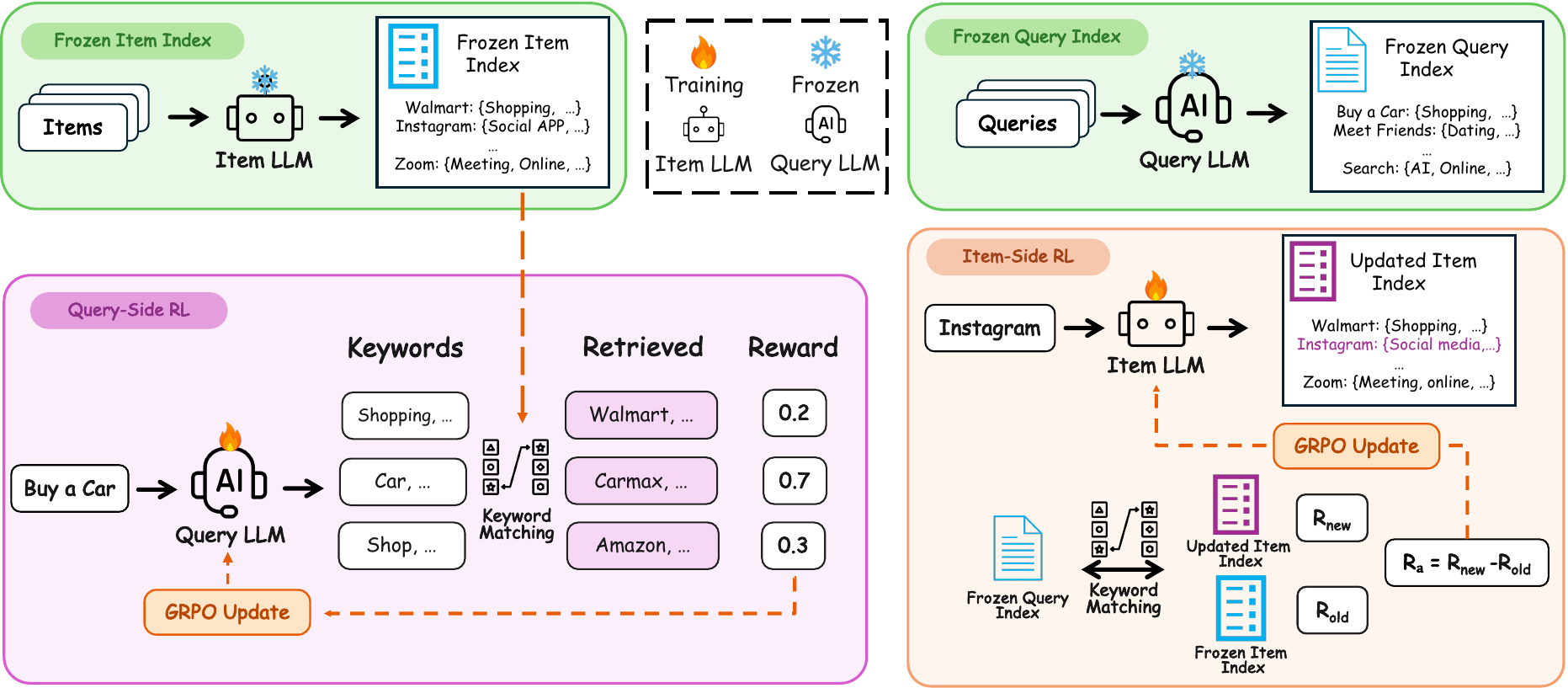}%
  }{%
    \fbox{\parbox[c][4.5cm][c]{0.995\linewidth}{\centering\sffamily\color{light}%
      \texttt{figure/main.pdf} --- overview figure goes here}}%
  }
  \caption{\textbf{Co-evolving reinforcement learning in \method{}.} Training alternates between query-side and item-side optimization, with the opposite-side index kept frozen during each stage. \emph{Left: Query-side RL.} The item LLM from the previous round constructs a frozen item index. For each query, the query LLM samples keyword sets, retrieves items through keyword matching, and is optimized with GRPO using the resulting retrieval $F_1$ as reward. \emph{Right: Item-side RL.} The updated query LLM constructs a frozen query index. For each sampled keyword set $S_i$, we construct a counterfactual item index by replacing only the reference keywords of item $i$ while leaving all other items unchanged. The item LLM is optimized with GRPO using the resulting marginal contribution to query-side retrieval quality, as defined in Equation \ref{eq:item_reward}.}
  \label{fig:overview}
\end{figure*}

\method{} contains two stages, as illustrated in \Cref{fig:overview}. First, a supervised fine-tuning (SFT) stage initializes both generators. Then, a reinforcement-learning stage alternately updates the query-side and item-side generators to optimize retrieval quality.

% In this work, we focus on lexical matching,  The ground truth is given by a \emph{gold relevant set} $\mathrm{rel}(q)$, while the system outputs a \emph{retrieved set} $I_{\mathrm{ret}}(q)$. A desirable retrieval system should include as many relevant items as possible while excluding irrelevant ones. We therefore evaluate the retrieved set against the gold set using standard metrics: precision $P$, recall $R$, and $F_1$:
% \begin{equation}
% P = \frac{|\mathrm{rel}(q) \cap I_{\mathrm{ret}}(q)|}{|I_{\mathrm{ret}}(q)|}, \qquad
% R = \frac{|\mathrm{rel}(q) \cap I_{\mathrm{ret}}(q)|}{|\mathrm{rel}(q)|}, \qquad
% F_1\bigl(I_{\mathrm{ret}}(q), \mathrm{rel}(q)\bigr) = \frac{2\, P\, R}{P + R}.
% \label{eq:f1}
% \end{equation}

% In industry practice, both precision and recall are closely tied to business outcomes. Precision reflects the system’s ability to filter out irrelevant items, which directly affects user experience. Recall measures the coverage of relevant candidates passed to downstream ranking and bidding stages, which is closely related to potential revenue. Therefore, we use the $F_1$ score as our primary focus, as it jointly accounts for precision and recall.

\subsection{Phase 1 --- Initialization with Supervised Fine-tuning}
\label{sec:sft}

Phase 1 of \method{} serves two purposes. First, it establishes an aligned keyword space between the query and item generators. Second, it provides RL with a meaningful initialization, ensuring sufficient initial recall to produce informative reward signals rather than requiring exploration from scratch.

\begin{wrapfloat}{algorithm}{l}{0.45\linewidth}
\caption{SFT initialization of both sides.}
\label{alg:sft}
\small

\begin{algorithmic}[1]
\Require Query and item sets $\mathcal{Q}, \mathcal{I}$, base LLM $G_0$, budgets $M,N$
\ForAll{items $i\in\mathcal{I}$}
    \State $S_i \gets M$ keywords sampled from $G_0(i)$
\EndFor
\ForAll{queries $q\in\mathcal{Q}$}
    \State $\mathcal{B}_q \gets \biguplus_{i\in\mathrm{rel}(q)} S_i$
    \Comment{multiset}
    \State $S_q \gets$ top-$N$ most frequent keywords of $\mathcal{B}_q$
\EndFor
\State $G^q_{\mathrm{SFT}} \gets\mathrm{SFT}(
G_0;
\{(q,S_q)\}_{q\in\mathcal{Q}}
)$

\State $G^i_{\mathrm{SFT}}
\gets
\mathrm{SFT}(
G_0;
\{(i,S_i)\}_{i\in\mathcal{I}}
)$
\end{algorithmic}
\end{wrapfloat}

The SFT data construction is summarized in \Cref{alg:sft}. We first use the original LLM to generate $M$ item-side keywords for each item $i$, obtaining an initial keyword set $S_i$. We then construct query-side targets using the relevance labels. For each query $q$, we collect relevant items, pool their initial item-side keywords, and select the top-$N$ most frequent keywords as the query-side target keyword set $S_q$. This construction ties each query-side keyword to keywords associated with its relevant items, thereby creating keyword overlap between relevant query--item pairs. Finally, we train the query-side generator $G^q$ on $\{(q, S_q)\}$ and the item-side generator $G^i$ on $\{(i, S_i)\}$ with SFT, yielding the initialized policies $G^q_{\mathrm{SFT}}$ and $G^i_{\mathrm{SFT}}$.

\subsection{Phase 2 --- Co-Evolving Reinforcement Learning}
\label{sec:rl}
After SFT, the RL stage further optimizes the two generators directly for retrieval quality. We adopt an alternating training paradigm: the query-side and item-side generators are updated in turn, while the inverted index constructed from the other side is kept frozen. This allows each generator to be optimized against a fixed retrieval environment rather than a moving target. In this section, we first introduce the specific reward design for each side, and then describe the alternating procedure that couples the two generators.

\paragraph{Query-Side RL.}
Given the item indexes generated by the frozen item-side LLM, query-side RL optimizes the query-side generator to improve retrieval quality. For each query $q \in \mathcal{Q}$, the query-side generator produces a keyword set $S_q$. The generated keywords are then matched against the item indexes to obtain the retrieved item set $I_{\mathrm{ret}}(q)$. Ideally, the retrieved set should cover as many relevant items in $\mathrm{rel}(q)$ while including as few irrelevant items as possible. We therefore use the $F_1$ score to evaluate retrieval quality:
\begin{equation}
P(q)=\frac{|\mathrm{rel}(q)\cap I_{\mathrm{ret}}(q)|}{|I_{\mathrm{ret}}(q)|}, \qquad R(q)=\frac{|\mathrm{rel}(q)\cap I_{\mathrm{ret}}(q)|}{|\mathrm{rel}(q)|}, \qquad F_1\bigl(I_{\mathrm{ret}}(q),\mathrm{rel}(q)\bigr)=\frac{2P(q)R(q)}{P(q)+R(q)}.
\label{eq:f1}
\end{equation}
In industrial retrieval systems, both precision and recall are closely tied to business outcomes. Precision reflects the system's ability to filter out irrelevant items and therefore directly affects the user experience. Recall measures the coverage of relevant candidates passed to downstream ranking and bidding stages and is thus closely related to potential revenue. The $F_1$ score balances these two objectives and serves as a good reward signal.

To constrain the number of generated keywords, we further impose a maximum keyword budget $K_{\max}$. A sampled keyword set receives a reward of $0$ if it exceeds the budget constraint:
\begin{equation}
\mathcal{R}_q(S_q)=\begin{cases}F_1\bigl(I_{\mathrm{ret}}(q),\mathrm{rel}(q)\bigr), & |S_q|\leq K_{\max}, \\ 0, & |S_q|>K_{\max}.\end{cases}
\label{eq:query_reward}
\end{equation}
Following the GRPO training paradigm \citep{shao2024deepseekmath}, we sample multiple keyword sets for each query and compute a terminal reward for each rollout. The rewards are then normalized within each rollout group to obtain relative advantage signals for optimizing the query-side generator.

\paragraph{Item-Side RL.}
Item-side RL aims to improve query-to-item retrieval by refining item keywords. Given a candidate keyword set $S_i$ for item $i$, we evaluate its quality by the change it induces in the overall retrieval quality.\footnote{We measure overall retrieval quality by summing the per-query $F_1$ scores defined in \eqref{eq:f1}.} This requires isolating the contribution of $S_i$ from those of all other items.

Specifically, at the beginning of each item-side update round, we freeze the query-side index and take the current item-side index as the reference state. We then construct a counterfactual index by replacing only the reference keywords of item $i$ with $S_i$, while leaving the keywords of all other items unchanged. For each query $q$, let $I_{\mathrm{ret}}^{\mathrm{ref}}(q)$ and $I_{\mathrm{ret}}^{\mathrm{cand}}(q;S_i)$ denote the retrieved item sets under the reference and counterfactual indexes, respectively. We define the item-side reward as
\begin{equation}
\mathcal{R}_i(S_i)=
\begin{cases}
\displaystyle
\underbrace{
\sum_{q\in\mathcal{Q}}
F_1\!\left(
I_{\mathrm{ret}}^{\mathrm{cand}}(q;S_i),
\mathrm{rel}(q)
\right)
}_{R_{\mathrm{new}}}
-
\underbrace{
\sum_{q\in\mathcal{Q}}
F_1\!\left(
I_{\mathrm{ret}}^{\mathrm{ref}}(q),
\mathrm{rel}(q)
\right)
}_{R_{\mathrm{old}}},
& \text{if } |S_i|\leq K_{\max},\\[4pt]
-1, & \text{otherwise}.
\end{cases}
\label{eq:item_reward}
\end{equation}

The first term in \eqref{eq:item_reward} measures the aggregate retrieval quality under the counterfactual index, while the second measures that under the reference index. Their difference therefore isolates the effect attributable to $S_i$. This difference-based formulation also enables efficient reward computation, as queries whose $F_1$ scores remain unchanged cancel out in \eqref{eq:item_reward}. Rather than running retrieval and computing $F_1$ for every query, we use the query-side inverted index together with cached query retrieval results to identify the affected queries and calculate only their contributions. We provide further implementation details in Appendix~\ref{app:examples}.

\begin{wrapfloat}{algorithm}{r}{0.43\linewidth}
\caption{Co-evolving RL loop.}
\label{alg:loop}
\small
\begin{algorithmic}[1]
\State \textbf{Initialize:} Build item index $\mathcal{I}^a_{0}$ from $G^a_{\mathrm{SFT}}$
\For{iteration $i = 1, 2, \dots$}
  \State Train $G^q_i$ against frozen item index $\mathcal{I}^a_{i-1}$
  \State Build query index $\mathcal{I}^q_i$ from $G^q_i$
  \State Train $G^a_i$ against frozen query index $\mathcal{I}^q_i$
  \State Build item index $\mathcal{I}^a_i$ from $G^a_i$
\EndFor
\end{algorithmic}
\end{wrapfloat}
\paragraph{Co-Evolving \& Discussion.}
Phase 2 of \method{} is an iterative approach between query-side and item-side RL starting from the post-SFT models. The first query-side RL phase uses the $G_{SFT}$ as index target. In later stages, each side is optimized against an index built by the latest model on the other side. This creates a co-evolution process in which the two generators progressively adapt to each other's keyword space and jointly evolve to improve the retrieval quality. The overall loop is summarized in \Cref{alg:loop}.

The framework also offers considerable flexibility. The same reward design can also be used with other RL algorithms such as PPO \citep{schulman2017proximal}. In applications where precision and recall have different business priorities, the standard $F_1$ reward can also be replaced by a weighted F-measure \citep{vanrijsbergen1979information}. Unless otherwise specified, we use the standard $F_1$ score throughout this work.

\section{Experiment Setting}
\label{sec:setting}

\subsection{Datasets}
\label{sec:datasets}
We evaluate our method on two industrial search datasets. The first is an internal APP marketplace search dataset consisting of de-identified, randomly sampled user queries. Each item corresponds to an application and is represented by its title and description. The second is WANDS \citep{chen2022wands}, a public product-search dataset from Wayfair, where each item corresponds to a product and is likewise represented by its title and description. Both datasets provide categorical relevance annotations for $(\text{query}, \text{item})$ pairs, which we binarize into relevant and irrelevant classes (see Appendix \ref{app:data} for details). We split the data only along the query dimension while retaining the full item universe for both training and validation, so that performance on validation set reflects generalization to unseen queries. Dataset statistics, including the number of relevant items per query, are summarized in Table \ref{tab:datasets}. We focus on these datasets because practical retrieval often involves many relevant items per query. In contrast, many conventional information retrieval datasets typically provide much sparser relevance annotations and thus less faithfully reflect this many-to-many setting.

\begin{table}[h]
\centering
\caption{Key statistics of our Internal APP marketplace search (Internal) dataset and WANDS dataset.}
\label{tab:datasets}
\small
\begin{tabular}{lccc}
\toprule
Dataset & Queries (Train / Evaluation) & Item universe & Relevant items per query \\
\midrule
Internal & $13{,}500$ / $1{,}500$ & $39{,}600$ Applications & ${\approx} 1,000$ \\
WANDS    & $430$ / $50$          & $42{,}994$ Products              & ${\approx} 200$ \\
\bottomrule
\end{tabular}
\end{table}

\subsection{Baseline Methods}
\label{sec:baselines}
We compare \method{} against three representative families of retrieval baselines, selecting widely used methods from each category. Additional implementation details are provided in Appendix \ref{app:training}.
\begin{itemize}
\item \textbf{Sparse (lexical) retrieval}, which scores query--item matches over sparse term representations. We include the classical BM25 \citep{robertson2009bm25} and the learned sparse retriever SPLADE-v2 \citep{formal2022spladev2}.

\item \textbf{Dense retrieval}, which maps queries and items into a shared embedding space and retrieves items based on nearest-neighbor similarity. We include DPR \citep{karpukhin2020dpr} and ANCE \citep{xiong2020approximate}. To provide a stronger dense baseline at a model scale comparable to \method{}, we additionally evaluate Qwen3-Embedding-4B \citep{qwen3} in both zero-shot and ANCE-finetuned settings.

\item \textbf{Generative retrieval}, which directly generates item identifiers using a sequence model. We include DSI \citep{tay2022dsi}, DSI-QG \citep{zhuang2022bridging}, and RIPOR \citep{zeng2024ripor}. We further include DeepRetrieval \citep{jiang2025deepretrieval}, which trains a query-rewriting language model with reinforcement learning and is therefore closely related to our query-side RL formulation.
\end{itemize}

\subsection{Training Setup}
\label{sec:training}
We instantiate two separate generators, one for the query side and one for the item side, using the same backbone architecture at each model scale: Qwen3-4B-Instruct for \method{}-4B and Qwen3-1.7B for \method{}-1.7B. Both Phase 1 and Phase 2 use the same prompt template shown in \Cref{fig:prompts}. We directly generate keywords without chain-of-thought prompting to avoid additional inference overhead. In Phase 1 (SFT), we set the query-side top-$N$ to $15$ and the item-side keyword budget $M$ to $10$. In Phase 2 (co-evolving RL), we train both generators with GRPO \citep{shao2024deepseekmath} using the \texttt{verl} framework \citep{sheng2024hybridflow}, with the generated keyword set capped at $K_{\max}=30$ on both sides. We alternate query- and item-side optimization for five rounds, training the query generator for $10$ GRPO epochs and the item generator for $5$ epochs per round. Unless otherwise specified, subsequent analyses use \method{}-4B. Additional training and implementation details are provided in Appendix \ref{app:training}.

\section{Experimental Results}
\label{sec:results}
\subsection{Main Results}
\label{sec:main_results}

The main results are reported in Table~\ref{tab:main}, with additional results at different retrieval cutoffs provided in Appendix~\ref{app:cutoffs}. Table~\ref{tab:main} reports precision, recall, and $F_1$, together with MRR, NDCG, and precision, recall, and $F_1$ at the top-100 cutoff, while Table~\ref{tab:cutoffs} provides corresponding results at cutoffs of 10 and 1000. For \method{}, retrieval rankings are obtained using BM25 over the generated keywords, as described in Section~\ref{sec:overview}. For the precision, recall, and $F_1$ scores of baseline methods, we sweep over retrieval cutoffs on the training set, select the cutoff that yields the best training $F_1$, and report the corresponding performance on the validation dataset. Overall, \method{} consistently achieves the highest $F_1$ score, demonstrating a clear advantage over strong baselines. The key observations of Table \ref{tab:main} are as follows:

\begin{itemize}[leftmargin=*]
\item \method{} achieves the strongest and most consistent performance across datasets, obtaining the best overall $F_1$ on both datasets, with scores of $0.396$ and $0.682$, respectively. Among the baselines, dense retrieval is comparatively more stable across the two datasets, with ANCE-Qwen4B consistently serving as the strongest baseline. In contrast, sparse retrieval performs substantially worse on the more challenging Internal dataset, while generative retrieval baselines degrade notably on the smaller and simpler WANDS dataset.

\item The co-evolving design of \method{} is necessary for strong retrieval performance. \method{} substantially outperforms both \method{}\frozen and DeepRetrieval, which optimize only the query-side keyword generator or query rewriter while keeping the item-side representations fixed. This consistent gap indicates that improving only the query-side representation is insufficient, and that jointly adapting the query and item keyword spaces is important for learning a well-aligned retrieval system.
\end{itemize}

\begin{table}[t]
\centering
\caption{Retrieval performance on the validation datasets. Columns suffixed with @100 (MRR@100, NDCG@100, P@100, R@100, $F_1$@100) are computed on the top-$100$ retrieved items, while $P$, $R$, and $F_1$ are the macro-averaged precision, recall, and $F_1$ over the full retrieved set. \faSnowflake\ denotes that the Item-side parameters are frozen. The best value in each column within a dataset is shown in \textbf{bold}. Metrics at cutoffs $10$ and $1000$ are reported in Appendix \ref{app:cutoffs}.}
\label{tab:main}
\small
\setlength{\tabcolsep}{4pt}
\resizebox{\textwidth}{!}{%
\begin{tabular}{lclccccc@{\hspace{2em}}ccc}
\toprule
Dataset & Type & Method & MRR@100 & NDCG@100 & P@100 & R@100 & $F_1$@100 & P & R & $F_1$ \\
\midrule
\multirow{14}{*}{{Internal}}
 & \multirow{2}{*}{Sparse} & BM25                    & 0.6728 & 0.2654 & 0.3630 & 0.0377 & 0.0611 & 0.2453 & 0.1141 & 0.1056 \\
 &                         & SPLADE-v2               & 0.5741 & 0.3365 & 0.4152 & 0.0588 & 0.0943 & 0.2706 & 0.4486 & 0.3019 \\
 \cmidrule(l){2-11}
 & \multirow{4}{*}{Dense}  & Qwen3-4B emb.           & 0.7255 & 0.4102 & 0.4436 & 0.0638 & 0.1022 & 0.2310 & 0.2689 & 0.2191 \\
 &                         & DPR                     & 0.5037 & 0.2860 & 0.3532 & 0.0504 & 0.0807 & 0.2408 & 0.4058 & 0.2700 \\
 &                         & ANCE                    & 0.6158 & 0.3866 & 0.4544 & 0.0636 & 0.1026 & 0.2671 & 0.4396 & 0.2971 \\
 &                         & ANCE-Qwen4B             & 0.7572 & 0.4879 & 0.5691 & 0.0818 & 0.1312 & 0.3756 & 0.4354 & 0.3575 \\
 \cmidrule(l){2-11}
 & \multirow{8}{*}{\shortstack[c]{Generative\\Retrieval}}
  & DSI                     & 0.6323 & 0.3393 & 0.4234 & 0.0581 & 0.0935 & 0.2882 & 0.4790 & 0.3220 \\
 & & DSI-QG                 & 0.6644 & 0.3560 & 0.4413 & 0.0599 & 0.0966 & 0.2901 & \textbf{0.4814} & 0.3241 \\
 & & RIPOR                  & 0.6719 & 0.4146 & 0.4975 & 0.0674 & 0.1100 & 0.3371 & 0.3793 & 0.3167 \\
 & & DeepRetrieval\,4B      & 0.7454 & 0.4437 & 0.5128 & 0.0715 & 0.1159 & 0.2907 & 0.3333 & 0.2750 \\
 & & \textbf{CoGR\frozen\,1.7B}      & 0.6450 & 0.3575 & 0.4310 & 0.0588 & 0.0958 & 0.3177 & 0.2289 & 0.2399 \\
 & & \textbf{CoGR\,1.7B}             & 0.7055 & 0.4456 & 0.5244 & 0.0739 & 0.1190 & 0.3523 & 0.4169 & 0.3527 \\
 & & \textbf{CoGR\frozen\,4B}        & 0.7020 & 0.3958 & 0.4694 & 0.0661 & 0.1066 & 0.3335 & 0.2459 & 0.2617 \\
 & & \textbf{CoGR\,4B}               & \textbf{0.7667} & \textbf{0.4930} & \textbf{0.5844} & \textbf{0.0844} & \textbf{0.1349} & \textbf{0.3976} & 0.4569 & \textbf{0.3963} \\
\midrule
\multirow{14}{*}{{WANDS}}
 & \multirow{2}{*}{Sparse} & BM25                    & 0.8575 & 0.7047 & 0.6508 & 0.4092 & 0.4263 & 0.4768 & 0.6149 & 0.4418 \\
 &                         & SPLADE-v2               & 0.9090 & 0.7872 & 0.7340 & \textbf{0.4506} & 0.4678 & 0.5914 & 0.6050 & 0.4903 \\
 \cmidrule(l){2-11}
 & \multirow{4}{*}{Dense}  & Qwen3-4B emb.           & 0.6471 & 0.4718 & 0.4270 & 0.2572 & 0.2599 & 0.3371 & 0.3408 & 0.2634 \\
 &                         & DPR                     & 0.8861 & 0.7259 & 0.6794 & 0.4071 & 0.4222 & 0.5612 & 0.5581 & 0.4520 \\
 &                         & ANCE                    & 0.8903 & 0.7333 & 0.6932 & 0.4000 & 0.4218 & 0.5731 & 0.5485 & 0.4554 \\
 &                         & ANCE-Qwen4B             & 0.9085 & \textbf{0.7955} & 0.7506 & 0.4498 & 0.4717 & 0.6136 & 0.6092 & 0.5012 \\
 \cmidrule(l){2-11}
 & \multirow{8}{*}{\shortstack[c]{Generative\\Retrieval}}
  & DSI                     & 0.5862 & 0.4706 & 0.4620 & 0.2396 & 0.2527 & 0.3994 & 0.3468 & 0.2890 \\
 & & DSI-QG                 & 0.8517 & 0.6594 & 0.6270 & 0.3659 & 0.3867 & 0.5051 & 0.5044 & 0.4023 \\
 & & RIPOR                  & 0.7905 & 0.5966 & 0.5704 & 0.3044 & 0.3283 & 0.4760 & 0.4205 & 0.3595 \\
 & & DeepRetrieval\,4B      & 0.8592 & 0.7029 & 0.6473 & 0.4072 & 0.4234 & 0.4802 & 0.6155 & 0.4431 \\
 & & \textbf{CoGR\frozen\,1.7B}      & 0.8332 & 0.4652 & 0.5945 & 0.2123 & 0.2686 & 0.5073 & 0.3751 & 0.3664 \\
 & & \textbf{CoGR\,1.7B}             & 0.8903 & 0.6701 & 0.8398 & 0.3339 & 0.4236 & 0.7325 & 0.5503 & 0.5685 \\
 & & \textbf{CoGR\frozen\,4B}        & 0.8111 & 0.5710 & 0.6494 & 0.2941 & 0.3554 & 0.5691 & 0.4696 & 0.4662 \\
 & & \textbf{CoGR\,4B}               & \textbf{0.9225} & 0.7146 & \textbf{0.8417} & 0.4059 & \textbf{0.4778} & \textbf{0.7940} & \textbf{0.6885} & \textbf{0.6819} \\
\bottomrule
\end{tabular}%
}
\end{table}

\subsection{Co-evolving Dynamics}
\label{sec:dynamics}

\begin{figure}[h!]
\centering
\includegraphics[width=\linewidth]{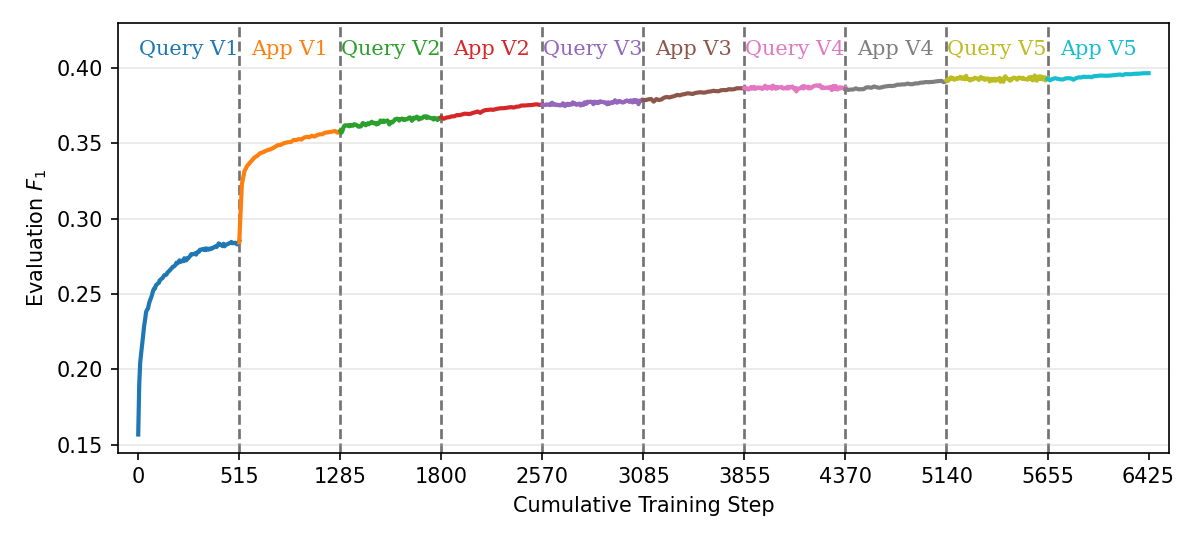}
\caption{Evaluation $F_1$ over cumulative training steps on the Internal dataset, starting from the SFT-initialized generators. Training alternates between query-side and item-side RL phases and labels $V1$--$V5$ denote the successive co-evolving rounds on each side. Evaluation $F_1$ increases from approximately $0.16$ before co-evolving RL to $0.40$ after five rounds of alternation.}
\label{fig:loop}
\end{figure}

\Cref{fig:loop} shows the validation $F_1$ over the alternating RL process, starting from the SFT-initialized generators. The metric increases approximately from $0.16$ before co-evolving to approximately $0.40$ after five rounds of co-evolving alternation. The largest gain occurs in the first round of training, where RL begins to align the two generated keyword spaces. Subsequent query- and item-side updates provide smaller but steady improvements.

\subsection{Ablation on Training Design}
\label{sec:ablation_training}

\begin{wraptable}{r}{0.45\linewidth}
\centering
\vspace{-\baselineskip}
\vspace{-1pt}
\caption{Ablation of training design choices on the Internal dataset. All experiments use Qwen3-4B as the base model.}
\label{tab:training_ablation}
\small
\setlength{\tabcolsep}{4pt}
\begin{tabular}{lccc}
\toprule
Variant & $P$ & $R$ & $F_1$ \\
\midrule
Transposed $F_1$  & 0.3482 & 0.4462 & 0.3743 \\
Shared generator  & 0.3678 & 0.4635 & 0.3798 \\
No SFT            & 0.3800 & 0.4283 & 0.3751 \\
\method{} (full)  & 0.3976 & 0.4569 & \textbf{0.3963} \\
\bottomrule
\end{tabular}
\vspace{-1pt}
\end{wraptable}

We ablate three design choices in \method{}: the marginal item-side reward in Equation~\ref{eq:item_reward}, the use of separate query- and item-side generators, and the SFT initialization in Phase~1. Specifically, \emph{Transposed $F_1$} replaces the marginal item-side reward with a symmetric item-centric retrieval objective. Analogous to the query-side reward in Equation~\ref{eq:query_reward}, we treat each item as a query and measure how well it retrieves its relevant queries:
\begin{equation*}
\mathcal{R}_i^{\mathrm{trans}}(S_i)=
\begin{cases}
F_1\bigl(Q_{\mathrm{ret}}(i), \mathrm{rel}(i)\bigr), & |S_i|\leq K_{\max},\\
0, & |S_i|>K_{\max},
\end{cases}
\label{eq:transposed_reward}
\end{equation*}
where $Q_{\mathrm{ret}}(i)$ denotes the set of queries retrieved for item $i$, and $\mathrm{rel}(i)$ denotes the set of queries relevant to $i$. \emph{Shared generator} uses a single generator for both query- and item-side keyword generation, so the co-evolving procedure alternately updates the same checkpoint for the two roles while maintaining separate reference policies. \emph{No SFT} removes the Phase~1 initialization and starts co-evolving RL directly from the base Qwen3-4B checkpoint.

As shown in \Cref{tab:training_ablation}, all three variants underperform the full \method{} model, supporting the effectiveness of each design choice. At the same time, all variants remain stable and achieve reasonable retrieval performance, suggesting that the overall co-evolving framework is robust to these alternative training configurations.

\subsection{Analysis of Keyword Evolution}
\label{sec:keyword_analysis}

\begin{figure*}[t]
\centering
\begin{subfigure}[c]{0.55\linewidth}
  \includegraphics[width=\linewidth]{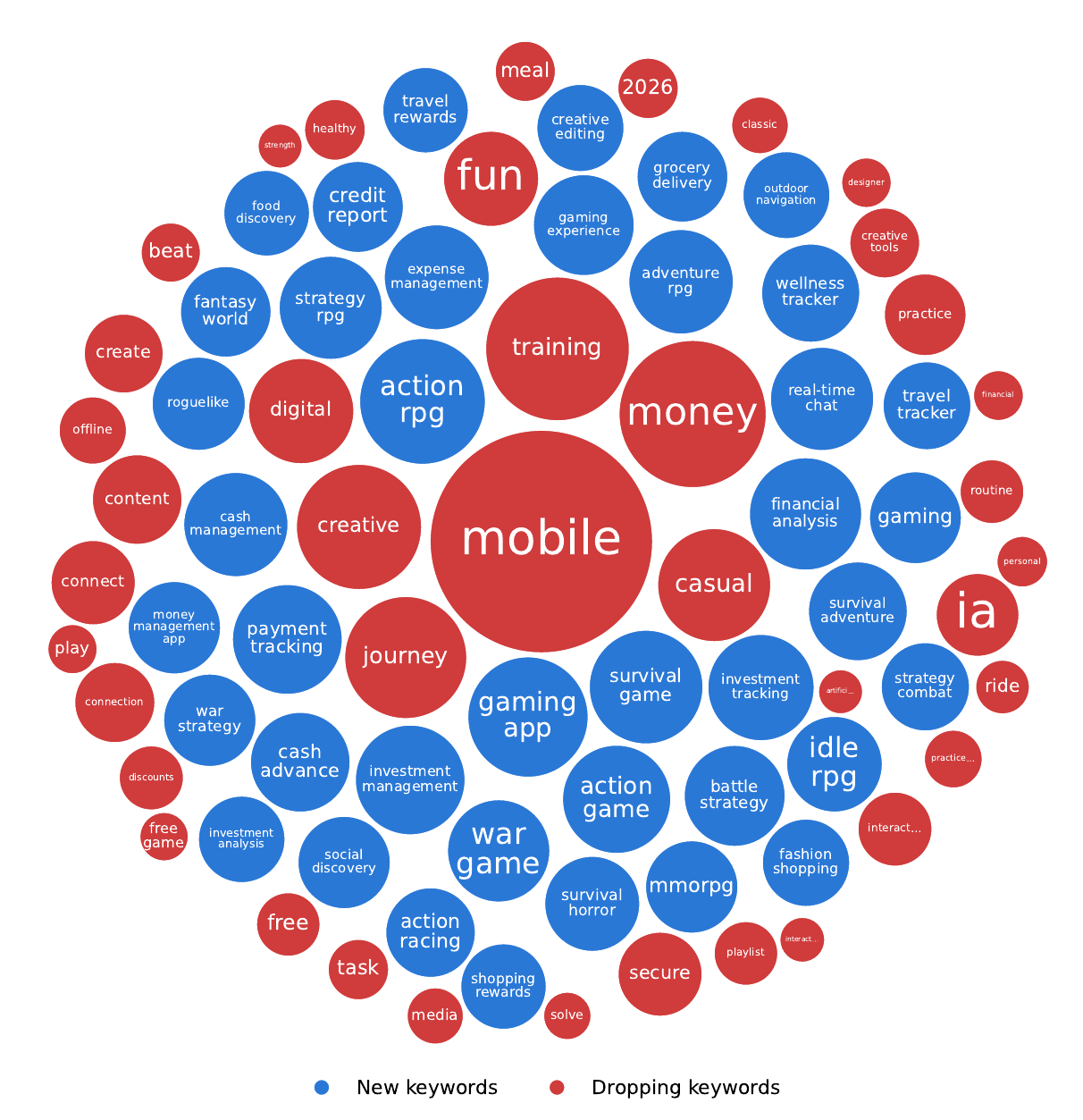}
  \caption{}
  \label{fig:kw_bubbles}
\end{subfigure}\hfill
\begin{minipage}[c]{0.44\linewidth}
  \begin{subfigure}[b]{\linewidth}
    \includegraphics[width=\linewidth]{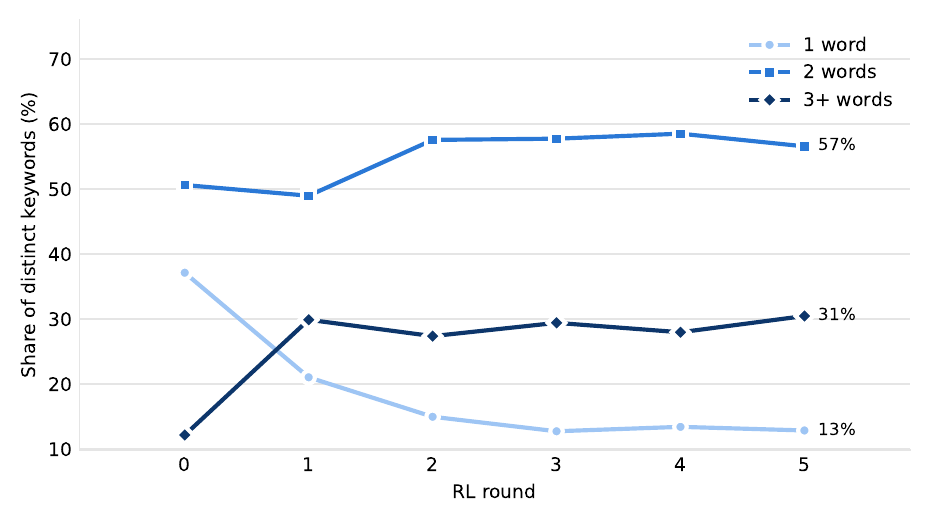}
    \caption{}
    \label{fig:kw_len}
  \end{subfigure}\\[4pt]
  \begin{subfigure}[b]{\linewidth}
    \includegraphics[width=\linewidth]{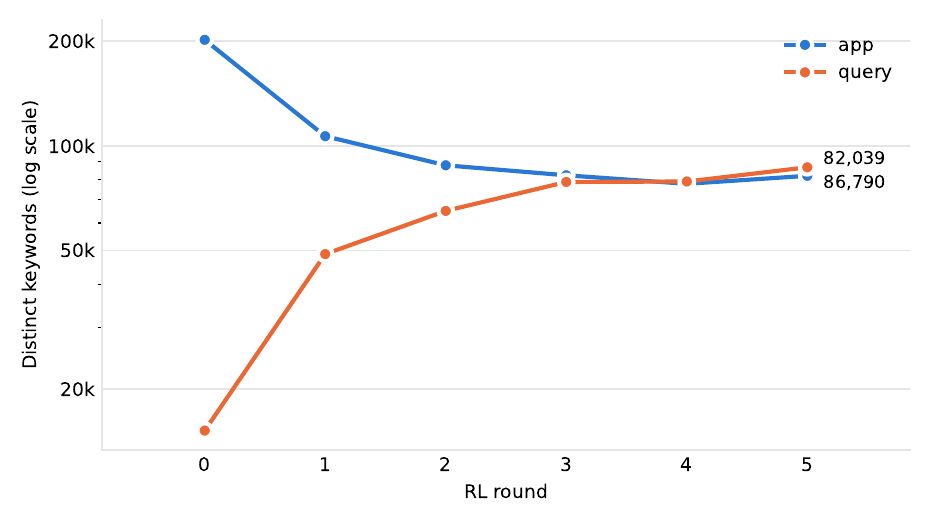}
    \caption{}
    \label{fig:kw_universe}
  \end{subfigure}
\end{minipage}
\caption{Keyword vocabulary dynamics under co-evolving RL on the Internal dataset. \emph{(a)} Keywords added (blue) and dropped (red) between the post-SFT vocabulary and that after five RL rounds; bubble area is proportional to the number of entities associated with each $n$-gram. \emph{(b)} Distribution of distinct $n$-grams by length across training rounds. (a) and (b) are computed over the union of query- and item-side keywords. \emph{(c)} Number of distinct indexed $n$-grams on the item (app) and query sides across rounds, shown on a log scale.}
\label{fig:keyword_dynamics}
\end{figure*}

In this section, we examine how the keyword space evolves during co-evolving RL. We compare the keyword distributions produced by the post-SFT generators with those obtained during five rounds of RL training as summarized in Figure~\ref{fig:keyword_dynamics}.

The first trend is that the learned keywords become increasingly specific. Co-evolving RL consistently reduces the prevalence of broad unigrams while increasing the use of more descriptive multi-word phrases. As shown in Figure~\ref{fig:kw_bubbles}, keywords that are removed during training are often generic terms such as ``mobile'' and ``fun'', whereas newly introduced keywords tend to be longer and more informative. This pattern is further supported by Figure~\ref{fig:kw_len}: the proportion of unigrams decreases from $37\%$ to $13\%$, while the proportion of phrases containing three or more words increases from $12\%$ to $31\%$.

We also find that the query- and item-side keyword spaces become more balanced over the course of training. As shown in Figure~\ref{fig:kw_universe}, the item-side vocabulary contracts during the early RL rounds, while the query-side vocabulary expands. The two keyword spaces then gradually converge to a similar number of unique keywords after approximately three to four rounds. This suggests that co-evolving RL not only refines the specificity of individual keywords, but also progressively aligns the query- and item-side vocabularies.

\subsection{Ablation on Additional Information}
\label{sec:ablation_info}
\begin{wraptable}{r}{0.45\linewidth}
\centering
\caption{Ablation on the information available to each generator. $-$~description leaves the item side with the title only, and $+$~search results additionally places search results in the query-side prompt. The experiments are performed on Internal data with Qwen3-4B-Instruct as the base model.}
\label{tab:info_ablation}
\small
\setlength{\tabcolsep}{4pt}
\begin{tabular}{lccc}
\toprule
Variant & $P$ & $R$ & $F_1$ \\
\midrule
\method{} (default) & 0.3976 & 0.4569 & \textbf{0.3963} \\
$-$ Description     & 0.3875 & 0.4204 & 0.3759 \\
$+$ Search Results  & 0.4381 & 0.5002 & 0.4379 \\
\bottomrule
\end{tabular}
\vspace{-15pt}
\end{wraptable}

We further study how retrieval quality depends on the textual information available to each generator. The default item-side input contains both the application title and description; we either remove the description or augment the query-side prompt with search results from the existing search system.

\Cref{tab:info_ablation} shows that richer context improves both sides of the system. Removing item descriptions reduces $F_1$ from $0.3963$ to $0.3759$, confirming that descriptions provide important lexical cues for items whose titles alone are underspecified. Conversely, adding search results substantially improves both precision and recall, raising $F_1$ to $0.4379$. As illustrated by the examples in \Cref{tab:examples}, the gain primarily comes from resolving ambiguous, misspelled, entity-centric, or non-English queries before keyword generation. These results suggest that \method{} benefits naturally from external semantic information.

\section{Related Works}
\label{sec:related}

\paragraph{Sparse and Dense Retrieval.}
Classical sparse retrieval uses lexical matching, such as TF--IDF and BM25 \citep{robertson2009bm25}, while learned sparse methods retain the inverted-index interface but learn term weights or lexical expansions, including docTTTTTquery \citep{nogueira2019doc2query}, DeepCT/DeepImpact \citep{dai2020deepct,mallia2021deepimpact}, uniCOIL \citep{lin2021unicoil}, and SPLADE \citep{formal2021splade,formal2022spladev2}. Dense retrieval follows the two-tower paradigm, with representative methods including DPR \citep{karpukhin2020dpr}, ANCE \citep{xiong2020approximate} with ANN-mined hard negatives, RocketQA \citep{qu2021rocketqa} with improved negative sampling and denoising, and GTR \citep{ni2022gtr} with scaled pretrained encoders. Sparse and dense retrieval remain widely used paradigms in large-scale retrieval systems.

\paragraph{Generative Retrieval.}
Generative retrieval replaces nearest-neighbor search with autoregressive generation of document identifiers. One line uses \emph{numeric identifiers}. DSI \citep{tay2022dsi,zhuang2022bridging} assigns documents semantic IDs and trains a sequence-to-sequence model to generate them from queries; NCI \citep{wang2022nci} improves this framework with query augmentation and prefix-aware decoding. Later methods learn or optimize identifiers rather than fixing them beforehand: GENRET \citep{sun2023genret} jointly learns document tokenization and retrieval, ASI \citep{yang2023auto} and MEVI \citep{zhang2023model} explore learned document indexing, while RIPOR \citep{zeng2024ripor} constructs identifiers from retrieval-oriented quantized representations and optimizes relevance across identifier prefixes.

A parallel line uses \emph{lexical identifiers} in the pretrained vocabulary space. GENRE \citep{cao2021genre} uses document titles, SEAL \citep{bevilacqua2022seal} uses document substrings with constrained generation, and MINDER \citep{li2023minder} combines multiple lexical views. More recent work moves beyond fixed lexical identifiers: GLEN \citep{lee2023glen} and NOVO \citep{wang2023novo} learn lexical identifiers from retrieval supervision, while ACID \citep{li2024acid} constructs abstractive identifiers such as generated keyphrases or summaries. Orthogonal to identifier design, prior work also improves generative retrieval through ranking- or relevance-aligned objectives \citep{li2024ltrgr,zhou2023genrrl,ddro2024,tang2024generative}, knowledge distillation \citep{roger2024}, and decoding strategies such as PAG \citep{zeng2024pag}.

\paragraph{LLMs and RL for Retrieval.} 
Recent work incorporates LLMs into retrieval by leveraging their generation capabilities to improve query representations. Prompting-based methods generate pseudo-documents, hypothetical answers, or rewritten queries that are subsequently consumed by sparse or dense retrievers \citep{gao2023hyde,wang2023query2doc,ma2023query,shen2024retrieval,lei2024corpus,liu2025real}. More recently, reinforcement learning with verifiable rewards (RLVR) has demonstrated strong empirical effectiveness across a range of domains \citep{deepseekr1,dai2026r1}, motivating its application to retrieval optimization. DeepRetrieval \citep{jiang2025deepretrieval} trains an LLM query generator directly from retrieval-metric rewards, while subsequent work extends this paradigm to downstream RAG utility, query reformulation, and retrieval-augmented search agents \citep{jiang2025s3,qin2025reinforced,ouyang2025token,jin2025search}. Related work further couples generation and retrieval by jointly training LLM-based query expansion and two-tower dense representations \citep{li2025reinforced,yao-etal-2025-expandr}.

\method{} instead focuses on co-evolving generative lexical retrieval, jointly training keyword generators on both the query and item sides. The generated keywords directly serve as the retrieval index. The two generators are alternately optimized toward a shared retrieval objective, allowing the lexical representations to co-adapt over training. This formulation is also related in spirit to recent advances in self-evolving LLM systems \citep{huang2026g, huang2025rzero}, where interacting roles improve against each other's evolving behavior.

\section{Conclusion}
\label{sec:conclusion}

We introduced \method{}, a co-evolving generative retrieval framework that trains LLMs to directly construct keyword-based retrieval representations on both query and item sides. By combining an aligned SFT initialization with alternating GRPO optimization against a frozen opposite-side index, \method{} enables the two keyword spaces to progressively co-adapt under a shared retrieval $F_1$ objective. Experiments on both an internal APP Marketplace dataset and WANDS demonstrate consistent gains over strong sparse, dense, and generative baselines.

More broadly, our work demonstrates the feasibility of a co-evolving keyword-matching framework in which query- and item-side representations are jointly adapted through retrieval feedback. This framework leaves several directions for future work. First, the reward can be extended beyond relevance metrics such as $F_1$ to downstream business objectives, including irrelevant ads percentage and revenue gain. Second, our current ranking stage uses BM25 over the generated keywords. Designing stronger retrieval ranking could further improve the overall retrieval quality.

\bibliographystyle{plainnat}
\bibliography{references}

%======================================================================
\appendix

\section{Relevance Label Details}
\label{app:data}

Both datasets provide categorical relevance labels for query–item pairs, which we convert into binary relevance labels for constructing the gold set. For the Internal dataset, labels are assigned by an internal LLM-as-a-judge on a five-level scale: \emph{excellent}, \emph{good}, \emph{acceptable}, \emph{poor}, and \emph{bad}. We treat pairs rated \emph{acceptable} or better as relevant, and all remaining pairs, including unlabelled pairs, as irrelevant. For WANDS \citep{chen2022wands}, human annotators provide three-level labels: \emph{Exact}, \emph{Partial}, and \emph{Irrelevant}. Following the same binary construction, we treat \emph{Exact} and \emph{Partial} as relevant, and treat \emph{Irrelevant} and unlabelled pairs as irrelevant.

\section{Training Details}
\label{app:training}
\subsection{Prompts}
The prompts used in phase 1 and 2 training are shown in Figure \ref{fig:prompts}.
\begin{figure}[h]
\centering
\begin{tcbraster}[raster columns=2, raster equal height=rows, raster column skip=3mm,
  colback=promptbg, colframe=promptframe, boxrule=0.8pt, arc=6pt,
  left=5pt, right=5pt, top=5pt, bottom=5pt, fontupper=\footnotesize]
\begin{tcolorbox}
\textbf{Query-side generator $G^q$}\par\smallskip
\textbf{System:} You expand user search queries into matching keywords.\par\smallskip
\textbf{User:} List keywords (synonyms, related terms, item categories) that the
query should match against. Output only a
comma-separated list of keywords, no explanation.\par\smallskip
User query: \ph{query}
\end{tcolorbox}
\begin{tcolorbox}
\textbf{Item-side generator $G^i$}\par\smallskip
\textbf{System:} You extract search keywords from item titles and descriptions.\par\smallskip
\textbf{User:} List keywords for lexical search matching. Output only a
comma-separated list of keywords, no explanation.\par\smallskip
Item title: \ph{item}\par\smallskip
Item Description: \ph{description}
\end{tcolorbox}
\end{tcbraster}
\caption{Prompts used to fine-tune the query-side ($G^q$) and item-side ($G^i$)
generators.}
\label{fig:prompts}
\end{figure}

\subsection{Hyperparameters}
We use the default sampling hyperparameters recommended for Qwen for Phase~1 Item keyword initialization (Section \ref{sec:sft}), as well as for model evaluation and index construction in Phase~2 (Section \ref{sec:rl}). During RL training, we instead use unconstrained stochastic sampling to encourage exploration. The decoding hyperparameters for each setting are summarized in Table~\ref{tab:sampling}. For GRPO training, we use 8 NVIDIA B200 GPUs. The key hyperparameters for query- and item-side optimization are summarized in Table~\ref{tab:rl-hparams}. We adopt a fully online RL setup, generating fresh rollouts from the current policy at each update to avoid off-policy training.

\begin{table}[h]
\centering
\caption{Decoding hyperparameters used for RL rollouts, initialization, evaluation, and index construction.}
\label{tab:sampling}
\begin{tabular}{lcccc}
\toprule
Setting & Temp. & top-$p$ & top-$k$ & min-$p$ \\
\midrule
RL rollout & $1.0$ & $1.0$ & $-1$ & $0$ \\
Initialization / Evaluation / Indexing & $0.7$ & $0.8$ & $20$ & $0$ \\
\bottomrule
\end{tabular}
\end{table}

\begin{table}[h]
\centering
\caption{GRPO hyperparameters for item-side and query-side training. Hyperparameter names follow the original \texttt{verl} configuration fields.}
\begin{tabular}{lcc}
\toprule
Hyperparameter & Item side & Query side \\
\midrule
Advantage estimator & GRPO & GRPO \\
\texttt{optim.lr} & $10^{-6}$ & $10^{-6}$ \\
\texttt{rollout.n} & $8$ & $8$ \\
\texttt{train\_batch\_size} & $256$ & $512$ \\
\texttt{ppo\_mini\_batch\_size} & $256$ & $512$ \\
\texttt{ppo\_micro\_batch\_size\_per\_gpu} & $32$ & $64$ \\
\texttt{n\_gpus\_per\_node} & $8$ & $8$ \\
\texttt{max\_response\_length} & $512$ & $512$ \\
\texttt{use\_kl\_loss} & False & False \\
Epochs & $5$ & $10$ \\
\bottomrule
\end{tabular}
\label{tab:rl-hparams}
\end{table}

\subsection{Efficient Implementation of Item-Side Reward}
\label{app:item_reward}

Naively computing the item-side reward in Eq.~\ref{eq:item_reward} would require rebuilding the item index and rerunning retrieval over all queries for every sampled item-side rollout. We avoid this cost by exploiting the fact that each counterfactual rollout changes the representation of only a single item.

At the beginning of each item-side training round, we cache the retrieval state under the frozen reference item index. For each query $q$, we store the number of retrieved items, the number of true positives, and the number of relevant items:
\begin{equation*}
n_q^{\mathrm{ret}} = \left|\mathcal{I}^{\mathrm{ref}}_{\mathrm{ret}}(q)\right|, \quad n_q^{\mathrm{tp}} = \left|\mathcal{I}^{\mathrm{ref}}_{\mathrm{ret}}(q) \cap \mathrm{rel}(q)\right|, \quad n_q^{\mathrm{rel}} = \left|\mathrm{rel}(q)\right|.
\end{equation*}
The reference $F_1$ score can then be computed directly from these cached counts:
\begin{equation*}
F_1^{\mathrm{ref}}(q) = \frac{2n_q^{\mathrm{tp}}}{n_q^{\mathrm{ret}} + n_q^{\mathrm{rel}}}.
\end{equation*}

We additionally cache, for each item $i$, the set of queries that retrieve it under its reference keyword set, denoted by $\mathcal{Q}^{\mathrm{ref}}_i$. Given a newly sampled keyword set $S_i$, we perform a single lookup against the frozen query-side inverted index to obtain the set of queries matched by the candidate item representation, denoted by $\mathcal{Q}^{\mathrm{cand}}_i$. Only queries for which the retrieval status of item $i$ changes can contribute to the reward. The affected query set is therefore
\begin{equation*}
\mathcal{Q}^{\Delta}_i = \mathcal{Q}^{\mathrm{cand}}_i \triangle \mathcal{Q}^{\mathrm{ref}}_i.
\end{equation*}

For each affected query $q$, replacing the keyword set of a single item can only add or remove item $i$ from the retrieved set. Its candidate $F_1$ score can therefore be updated directly from the cached counts:
\begin{equation*}
F_1^{\mathrm{cand}}(q) =
\begin{cases}
\displaystyle \frac{2\left(n_q^{\mathrm{tp}} + \mathbb{I}[i \in \mathrm{rel}(q)]\right)}{n_q^{\mathrm{ret}} + 1 + n_q^{\mathrm{rel}}}, & q \in \mathcal{Q}^{\mathrm{cand}}_i \setminus \mathcal{Q}^{\mathrm{ref}}_i, \\
\displaystyle \frac{2\left(n_q^{\mathrm{tp}} - \mathbb{I}[i \in \mathrm{rel}(q)]\right)}{n_q^{\mathrm{ret}} - 1 + n_q^{\mathrm{rel}}}, & q \in \mathcal{Q}^{\mathrm{ref}}_i \setminus \mathcal{Q}^{\mathrm{cand}}_i.
\end{cases}
\end{equation*}

All unaffected queries have identical reference and candidate $F_1$ scores and therefore cancel in the reference-subtracted objective. The item-side reward can thus be computed only over the affected queries:
\begin{equation*}
R_i(S_i) = \sum_{q \in \mathcal{Q}^{\Delta}_i} \left(F_1^{\mathrm{cand}}(q) - F_1^{\mathrm{ref}}(q)\right).
\end{equation*}

As a result, each sampled item-side rollout requires only one lookup against the frozen query-side index, followed by constant-time $F_1$ updates for the affected queries. This avoids rebuilding the full counterfactual index or rerunning query-to-item retrieval over the entire query set for every rollout.

\section{Retrieval Metrics at Additional Cutoffs}
\label{app:cutoffs}

\begin{table}[h]
\centering
\caption{Held-out query-side retrieval metrics at cutoffs $10$ and $1000$, using the same method grouping as \Cref{tab:main}. \faSnowflake\ denotes that the Item-side parameters are frozen. The best value in each column within a dataset is shown in \textbf{bold}.}
\label{tab:cutoffs}
\small
\setlength{\tabcolsep}{4pt}
\resizebox{\textwidth}{!}{%
\begin{tabular}{lclccccc@{\hspace{2em}}ccccc}
\toprule
Dataset & Type & Method & MRR@10 & NDCG@10 & P@10 & R@10 & $F_1$@10 & MRR@1000 & NDCG@1000 & P@1000 & R@1000 & $F_1$@1000 \\
\midrule
\multirow{14}{*}{{Internal}}
 & \multirow{2}{*}{Sparse} & BM25                    & 0.6682 & 0.3281 & 0.4815 & 0.0066 & 0.0128 & 0.6729 & 0.1640 & 0.2453 & 0.1141 & 0.1056 \\
 &                         & SPLADE-v2               & 0.5683 & 0.2660 & 0.4471 & 0.0069 & 0.0133 & 0.5744 & 0.3938 & 0.3054 & 0.3523 & 0.2895 \\
 \cmidrule(l){2-13}
 & \multirow{4}{*}{Dense}  & Qwen3-4B emb.           & 0.7212 & 0.4185 & 0.5653 & 0.0091 & 0.0176 & 0.7256 & 0.3339 & 0.2310 & 0.2689 & 0.2191 \\
 &                         & DPR                     & 0.4955 & 0.2250 & 0.3793 & 0.0057 & 0.0110 & 0.5041 & 0.3466 & 0.2677 & 0.3132 & 0.2551 \\
 &                         & ANCE                    & 0.6109 & 0.3348 & 0.5156 & 0.0079 & 0.0153 & 0.6161 & 0.4023 & 0.3062 & 0.3490 & 0.2888 \\
 &                         & ANCE-Qwen4B             & 0.7544 & \textbf{0.4261} & 0.6445 & 0.0104 & 0.0201 & 0.7573 & \textbf{0.4989} & 0.3756 & \textbf{0.4354} & 0.3575 \\
 \cmidrule(l){2-13}
 & \multirow{8}{*}{\shortstack[c]{Generative\\Retrieval}}
  & DSI                     & 0.6265 & 0.2838 & 0.4846 & 0.0069 & 0.0134 & 0.6325 & 0.4113 & 0.3201 & 0.3738 & 0.3053 \\
 & & DSI-QG                 & 0.6597 & 0.3020 & 0.5121 & 0.0073 & 0.0141 & 0.6647 & 0.4172 & 0.3234 & 0.3753 & 0.3077 \\
 & & RIPOR                  & 0.6671 & 0.3460 & 0.5527 & 0.0079 & 0.0155 & 0.6722 & 0.4370 & 0.3371 & 0.3793 & 0.3167 \\
 & & DeepRetrieval\,4B      & 0.7422 & 0.4044 & 0.6154 & 0.0097 & 0.0187 & 0.7456 & 0.3992 & 0.2907 & 0.3333 & 0.2750 \\
 & & \textbf{CoGR\frozen\,1.7B}      & 0.6402 & 0.3111 & 0.5046 & 0.0076 & 0.0147 & 0.6452 & 0.2747 & 0.3251 & 0.2068 & 0.2302 \\
 & & \textbf{CoGR\,1.7B}             & 0.7019 & 0.3888 & 0.5966 & 0.0092 & 0.0177 & 0.7057 & 0.4146 & 0.3806 & 0.3408 & 0.3296 \\
 & & \textbf{CoGR\frozen\,4B}        & 0.6990 & 0.3578 & 0.5695 & 0.0090 & 0.0173 & 0.7022 & 0.2985 & 0.3431 & 0.2232 & 0.2517 \\
 & & \textbf{CoGR\,4B}               & \textbf{0.7643} & 0.4198 & \textbf{0.6576} & \textbf{0.0106} & \textbf{0.0202} & \textbf{0.7668} & 0.4621 & \textbf{0.4269} & 0.3810 & \textbf{0.3716} \\
\midrule
\multirow{14}{*}{{WANDS}}
 & \multirow{2}{*}{Sparse} & BM25                    & 0.8556 & 0.7314 & 0.7816 & 0.0564 & 0.1006 & 0.8575 & 0.7907 & 0.3149 & 0.8330 & 0.3806 \\
 &                         & SPLADE-v2               & 0.9090 & \textbf{0.7861} & \textbf{0.8780} & 0.0636 & \textbf{0.1125} & 0.9091 & 0.8530 & 0.2732 & 0.8818 & 0.3384 \\
 \cmidrule(l){2-13}
 & \multirow{4}{*}{Dense}  & Qwen3-4B emb.           & 0.6442 & 0.5129 & 0.5780 & 0.0456 & 0.0800 & 0.6476 & 0.5478 & 0.1759 & 0.5915 & 0.2150 \\
 &                         & DPR                     & 0.8850 & 0.7539 & 0.8460 & 0.0631 & 0.1113 & 0.8861 & 0.8199 & 0.2692 & 0.8662 & 0.3318 \\
 &                         & ANCE                    & 0.8900 & 0.7617 & 0.8580 & \textbf{0.0640} & \textbf{0.1125} & 0.8905 & 0.8133 & 0.2726 & 0.8467 & 0.3334 \\
 &                         & ANCE-Qwen4B             & 0.9067 & 0.7816 & 0.8620 & 0.0603 & 0.1072 & 0.9085 & \textbf{0.8591} & 0.2785 & \textbf{0.8883} & 0.3447 \\
 \cmidrule(l){2-13}
 & \multirow{8}{*}{\shortstack[c]{Generative\\Retrieval}}
  & DSI                     & 0.5814 & 0.4717 & 0.5580 & 0.0343 & 0.0606 & 0.5867 & 0.5782 & 0.2244 & 0.6451 & 0.2632 \\
 & & DSI-QG                 & 0.8517 & 0.6827 & 0.7800 & 0.0548 & 0.0972 & 0.8521 & 0.7474 & 0.2544 & 0.7891 & 0.3067 \\
 & & RIPOR                  & 0.7902 & 0.6476 & 0.7500 & 0.0514 & 0.0907 & 0.7907 & 0.6857 & 0.2416 & 0.7259 & 0.2889 \\
 & & DeepRetrieval\,4B      & 0.8550 & 0.7320 & 0.7857 & 0.0570 & 0.1016 & 0.8592 & 0.7956 & 0.3280 & 0.8354 & 0.3915 \\
 & & \textbf{CoGR\frozen\,1.7B}      & 0.8315 & 0.6089 & 0.7236 & 0.0468 & 0.0841 & 0.8333 & 0.4293 & 0.5336 & 0.3713 & 0.3726 \\
 & & \textbf{CoGR\,1.7B}             & 0.8903 & 0.7427 & 0.8391 & 0.0589 & 0.1050 & 0.8903 & 0.6188 & 0.7787 & 0.5515 & 0.5814 \\
 & & \textbf{CoGR\frozen\,4B}        & 0.8079 & 0.6462 & 0.7408 & 0.0527 & 0.0941 & 0.8111 & 0.5184 & 0.5848 & 0.4736 & 0.4732 \\
 & & \textbf{CoGR\,4B}               & \textbf{0.9209} & 0.7537 & 0.8655 & 0.0595 & 0.1062 & \textbf{0.9225} & 0.7075 & \textbf{0.8115} & 0.6633 & \textbf{0.6703} \\
\bottomrule
\end{tabular}%
}
\end{table}

\section{Example Generated Keywords}
\label{app:examples}

\Cref{tab:examples} shows how the generated keywords change over the alternating rounds,
and how they change again when each side is given additional information beyond the query
string or the item title: online search hints on the query side
(\Cref{tab:query_examples}) and the application description on the item side
(\Cref{tab:app_examples}).

\begin{table}[h]
\centering
\caption{Example generated keywords before and after co-evolving RL, and with additional
information at the input.}
\label{tab:examples}
\begin{subtable}[t]{0.48\linewidth}
\centering
\caption{Query side. Round-$5$ keywords are also shown for a prompt that additionally
contains online search hints for the query.}
\label{tab:query_examples}
\footnotesize
\setlength{\tabcolsep}{3pt}
\renewcommand{\arraystretch}{1.2}
\begin{tabularx}{\linewidth}{l >{\raggedright\arraybackslash}X >{\raggedright\arraybackslash}X}
\toprule
Query & w/o Search & w/ Search \\
\midrule
\emph{family search} & album, anniversary, birthday, birthdays, digital, photo, search, sharing, tracker, ... & genealogy, family history, genealogy app, family calendar, ancestry, family tree, ... \\
\emph{\cjk{解压软件}} & relaxation, mindfulness, meditation, relaxation app, stress relief, puzzle game, mental health, ... & archive, compress, compression, extract, extractor, file, files, manager, rar, unzip, zip, ...  \\
\emph{zenless} &  mindfulness, meditation, relaxation, meditation app, mental health, mindfulness app, stress relief, ... & action, adventure, battle, combat, defense, fantasy, royale, rpg, ... \\
\emph{duch bros} & action, multiplayer, action game, arcade game, adventure game, action shooter, arcade, action rpg, ... & burger, coffee, deals, delivery, drink, fast, food, local, order, pizza, ... \\
\bottomrule
\end{tabularx}
\end{subtable}\hfill
\begin{subtable}[t]{0.48\linewidth}
\centering
\caption{Item side. Round-$5$ keywords are also shown for a prompt that additionally
contains the application description.}
\label{tab:app_examples}
\footnotesize
\setlength{\tabcolsep}{3pt}
\renewcommand{\arraystretch}{1.2}
\begin{tabularx}{\linewidth}{l >{\raggedright\arraybackslash}X >{\raggedright\arraybackslash}X}
\toprule
Application & w/o Desc. & w Desc. \\
\midrule
\emph{10000000} & 10000000, 1000 & rpg, dungeon crawler, puzzle, action rpg, roguelike, adventure,\dots \\
\emph{Compound Quest} & crypto, decentralized finance, dapp, yield farming, blockchain & compound words, word puzzles, vocabulary building, spelling skills, reading skills, esl learning, picture match \\
\emph{Ultra Plunge Tracker} & dive, dive log, diving, swim & cold plunge, sauna, steam room, ice bath, heart rate, recovery, wellness, ai coach, hyperbaric chamber \\
\emph{Life is Slow} & life is slow, slow life & nature, simulation, nature simulation, puzzle, relaxation, nature theme, adventure, nature game, nature therapy \\
\bottomrule
\end{tabularx}
\end{subtable}
\end{table}

\section{Baseline Methods}
\label{app:baselines}

\begin{wraptable}{r}{0.4\linewidth}
\centering
\vspace{-15pt}
\caption{Base models used for each baseline.}
\label{tab:baseline-models}
\small
\resizebox{\linewidth}{!}{
\begin{tabular}{ll}
\toprule
Baseline & Base model \\
\midrule
BM25            & --- \\
\midrule
DPR             & \texttt{bert-base-multilingual-uncased} \\
ANCE            & \texttt{roberta-base} \\
SPLADE-v2       & \texttt{naver/splade\_v2\_max} \\
Qwen3-4B emb. & \texttt{Qwen3-Embedding-4B} \\
ANCE-Qwen4B      & \texttt{Qwen3-Embedding-4B} \\
\midrule
DSI             & \texttt{mT5-base} \\
DSI-QG          & \texttt{mT5-base} \\
RIPOR           & \texttt{mT5-base} \\
DeepRetrieval   & \texttt{Qwen3-4B-Instruct-2507} \\
\bottomrule
\end{tabular}
}
\vspace{-20pt}
\end{wraptable}

We compare against three families of retrieval baselines, with the base model used for each method summarized in Table~\ref{tab:baseline-models}. For all trainable baselines, we follow their standard training setups and use the same query- and item-side information available to CoGR whenever applicable, including the same item title and description fields and aligned prompting inputs. This ensures that performance differences are not driven by additional textual information or task-specific prompt engineering.

For unconstrained precision, recall, and $F_1$, we tune the retrieval cutoff on the training split by selecting the value that maximizes training $F_1$, and then apply this fixed cutoff to the validation dataset. Checkpoints are selected according to their validation $F_1$ under the same protocol. Unless otherwise specified, trained baselines use 8 GPUs with distributed data parallelism and otherwise follow their standard optimization configurations.

\end{document}

%% file: apple_preamble.tex
\usepackage{amsmath}
\usepackage{enumerate}
\usepackage{algorithm}
\usepackage{algpseudocode}
\usepackage{amsfonts}
\usepackage{amsthm}
\usepackage{cleveref}
\usepackage{diagbox}
\usepackage{colortbl}
\usepackage{amssymb}
\usepackage{xspace}
\usepackage{wrapfig}
\usepackage{adjustbox}
\usepackage{tabularx}
\usepackage{booktabs}
\usepackage{mathtools}
\usepackage{tikz}
\usepackage{enumitem}
\usepackage{silence}
\usepackage{dsfont}
\usepackage[table]{xcolor}
\usepackage[dvipsnames]{xcolor}
\usepackage{multirow}
\usepackage{makecell}
\usepackage{xfakebold}
\input{math_commands}

\definecolor{textgray}{HTML}{6E6E73}
\usetikzlibrary{positioning, calc}
\usetikzlibrary{decorations.pathmorphing}

\makeatletter
\patchcmd{\wrong@fontshape}{\@gobbletwo}{}{}{}
\makeatother
\numberwithin{equation}{section}
\makeatletter
\AtBeginDocument{
  \urlstyle{sf}
  
}
\makeatother

\definecolor{light}{RGB}{125, 125, 125}
\crefname{tcb@cnt@pbox}{code}{code}
\Crefname{tcb@cnt@pbox}{Code}{Code}
\crefname{assumption}{assumption}{assumption}
\Crefname{assumption}{Assumption}{Assumptions}

\newtcolorbox[auto counter]{pbox}[2][]{
  colback=white,
  title=Code~\thetcbcounter: #2,
  #1,fonttitle=\sffamily,
  fontupper=\sffamily,
  arc=2pt,
  colframe=bgcolor,
  coltitle=fgcolor,
  colbacktitle=bgcolor,
  toptitle=0.25cm,
  bottomtitle=0.125cm
}

\makeatletter
\newcommand\applefootnote[1]{%
  \begingroup
  \renewcommand\thefootnote{}%
  \renewcommand\@makefntext[1]{\noindent##1}%
  \footnote{#1}%
  \addtocounter{footnote}{-1}%
  \endgroup
}
\makeatother

\definecolor{cverbbg}{gray}{0.90}

%% file: math_commands.tex
\usepackage{amsmath,amsfonts,bm}

\def\eqref#1{equation~\ref{#1}}
\def\1{\bm{1}}

\DeclareMathAlphabet{\mathsfit}{\encodingdefault}{\sfdefault}{m}{sl}
\SetMathAlphabet{\mathsfit}{bold}{\encodingdefault}{\sfdefault}{bx}{n}